\documentclass[twocolumn,pre,aps,superscriptaddress,a4paper]{revtex4}
\usepackage[dvips]{graphicx}
\usepackage[utf8]{inputenc}
\usepackage[T1]{fontenc}
\usepackage{lipsum}
\usepackage{float}
\usepackage[dvipsnames]{xcolor}
\usepackage{amsmath}
\usepackage{soul}
\usepackage{graphicx}
\usepackage{epsfig,amsmath,amsfonts,amssymb,
color,
}
\usepackage{siunitx}
\usepackage{comment}
\begin{document}

\title{
The vanishing water/oil interface in the presence of antagonistic salt
}

%
%
\author{Gudrun Glende}
\affiliation{Department of Mechanical and Industrial Engineering, Norwegian University of Science and Technology, 7491 Trondheim, Norway}

\author{Astrid S.\ de Wijn  }
\affiliation{Department of Mechanical and Industrial Engineering, Norwegian University of Science and Technology, 7491 Trondheim, Norway}
\affiliation{Department of Physics, Stockholm University, Sweden}

\author{Faezeh Pousaneh$^{\scriptscriptstyle{}}$\footnote{Corresponding author. Electronic address: \texttt{faezeh.pousaneh@ntnu.no}.}}  
\affiliation{Department of Mechanical and Industrial Engineering, Norwegian University of Science and Technology, 7491 Trondheim, Norway}

\begin{abstract}
Antagonistic salts are salts which consist of  hydrophilic and hydrophobic
ions. In a binary mixture of water and organic solvent, these ions
preferentially dissolve into different phases. We investigate the effect of
an antagonistic salt, tetraphenylphosphonium chloride PPh$_4$Cl, in a mixture
of water and 2,6-lutidine by means of Molecular Dynamics (MD) Simulations.  For
increasing concentrations of the salt the two-phase region is shrunk and the
interfacial tension in reduced, in contrast to what happens when a normal salt
is added to such a mixture.  The MD simulations allow us to investigate in
detail the mechanism behind the reduction of the surface tension.  We obtain
the ion and composition distributions around the interface and determine the
hydrogen bonds in the system and conclude that the addition of salt alter the
hydrogen bonding.
\end{abstract}
\maketitle

\section{Introduction}

Mixtures of water and oil are ubiquitous in nature and technology.  Biological systems are largely water-based, but also contain oily molecules.
In maritime and other off-shore applications, lubricants and other oils come into contact with seawater often.
What many of these mixtures have in common is the additional presence of salt ions.
This addition of a small amount of salt can significantly change the properties of an oil/water mixture.

Most binary mixtures of water and organic molecules are phase-separated at ambient conditions.
When a simple inorganic salt is added to the mixture, the two-phase region enlarges \cite{Balevicius:99:00}.
This picture changes for more complex salts with organic groups that are hydrophobic.
When an antagonistic (or amphiphilic) salt is added to an oil/water mixture this can lead to reduction of interfacial tension between water and oil, and therefore to a shrinkage of the two phase region.
It has been shown experimentally that the interfacial tension decreases with increasing the amount of antagonistic salts \cite{Luo:06:00}.
If enough salt is added, it can even make the interface disappear altogether and cause the oil and water to mix~\cite{sadakane:13:0,sadakane:11:0,sadakane:09:0,leys:13:0,sadakane:12:0}.
In addition to changes in the interfacial tension, adding an antagonistic salt can also lead to interesting structural changes, such as a lamellar phases  \cite{sadakane:13:0,sadakane:09:0,leys:13:0}.
These effects have been studied and explained theoretically for small salt concentration \cite{onuki:04:0} and for higher salt concentrations \cite{pousaneh:14:01,onuki:11:1,onuki:11:0}.
Analytical calculations of the ion distribution have suggested that the cations and anions of an antagonistic salt tend to adsorb around a water-oil interface  \cite{Onuki:06:00}.

Here we study this problem in a different way, using detailed atomistic Molecular Dynamics (MD) simulations.
Because such simulations give us the trajectories of all the particles in the system, they can be used as numerical experiments.
They allow us to observe in much more detail where the ions are and how they are behaving, which enables us to study and verify the mechanisms of the reduction in the interfacial tension.
We investigate a mixture of water and 2,6-lutidine (2,6-dimethylpyridin) with various concentrations of the antagonistic salt tetraphenylphosphonium chloride PPh$_4^+$Cl$^-$, with  hydrophilic anion Cl$^-$ and (PPh$_4^+$) anion.

The phase diagram and mixing behaviour of neat water/2,6-lutidine mixtures have  gained considerable attention in the context of colloidal systems \citep{Broide:93:00,Frisken:91:00,gallager:92:0}  and critical Casimir forces~\citep{gambassi:09:0,hertlein:08:0}. 
Critical Casimir forces rises between  two selective (hydrophilic or  hydrophobic) colloids in a suspension  when the liquid base of the suspension is near its critical point ~\citep{Fisher:78:0,krech:99:0}. 
 When colloids  are charged the force is tuned by electrostatic interactions~\citep{gambassi:09:0,hertlein:08:0,nellen:09:00,nellen:11:0,bier:11:0,pousaneh:11:00,pousaneh:12:00,pousaneh:14:00}. Moreover, the addition of a salt to the water/2,6-lutedine mixture allows one to tune these interactions and create different structures~\citep{sadakane:13:0,sadakane:09:0,leys:13:0,pousaneh:14:01}. Antagonistic salts are of particular interest in this context, because the two ions have substantially different interactions with the oil and water.

The neat water/2,6-lutidine binary mixture and its phase diagram have been studied experimentally and analytically \citep{Andon:52:00,Jayalakshmi:93:00} and recently by molecular dynamics simulations \citep{pousaneh:16:00}.                              
Without the additional salt, this mixture has a closed-loop phase diagram with a relatively wide temperature miscibility gap, which makes it very suitable for studying (de)mixing.
The lower critical point is close to room temperature, around  $T_c\approx  307.1$~K with the lutidine mole fraction $x_{lut}\approx 6.1\%$ \cite{Andon:52:00}. 

\begin{figure}[ht]
\centering 
\includegraphics[width=0.5\textwidth]{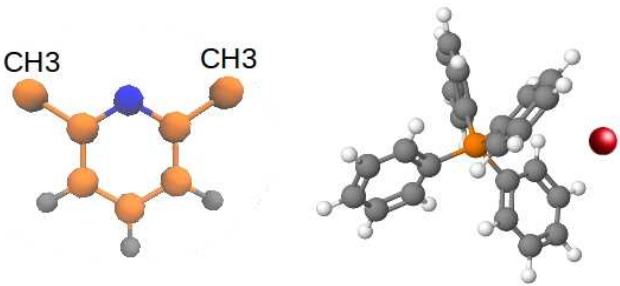} \vspace{-1cm}
\caption{(Left) Schematic representation of the 2,6-lutidine molecule, C$_7$H$_9$N. CH$_3$ groups are shown as single united atoms. For the charge distributions see ref.~\cite{pousaneh:16:00}. (Right) Schematic representation of salt  Ph$_4^+$Cl$^-$. Cation is  tetraphenylphosphonium PPh$_4^+$ with four phenyl rings, where the  orange atom in the middle represents the phosphonium. The anion is  Cl$^-$.}
\label{fig:1WEO}
\end{figure}

An antagonistic salt has one hydrophilic ion and one hydrophobic ion.
This makes them potentially very different from more simple hydrophilic salts, where both cation and anion are hydrophilic.
With the addition of such hydrophilic salts, the solubility of water and the organic solvent decreases and therefore the two-phase region is enhanced.
Rather than both ions preferring to dissolve in water, in antagonistic salts, one ion will prefer to dissolve in the water, while the other prefers to dissolve in the oil.
The effect of the
antagonistic salt Na$^+$BPh$_4^-$ on a binary mixture of D$_2$O and 3-methylpyridine (3MP) was recently studied experimentally by Sadakane  {\it {et al.}} \citep{sadakane:13:0,sadakane:09:0}
by means of small-angle neutron scattering and optical microscopy. They show that for increasing 
salt concentration the two-phase region shrinks and eventually even disappears.
A similar effect has been reported with the salt PPh$_4^+$Cl$^-$~\citep{sadakane:12:0} (see Fig.~\ref{fig:1WEO} (right)) that we study in this work. 
The hydrophobic cation has four phenyl rings that interfere with hydrogen bonding and the hydration shell.
Thus, the cations preferentially dissolve in an organic solvent (oil) whereas the anions prefer to stay within the water.
Consequently, the cations and anions behave antagonistically and may distribute heterogeneously when added to a binary liquid consisting of water and an organic solvent.
   
In this article,  we study the effect of
an antagonistic salt, tetraphenylphosphonium chloride PPh$_4$Cl, in a mixture
of water and 2,6-lutidine by means of atomistic MD Simulations. We investigate the addition of salt on properties of the mixture, such as surface tension and hydrogen bonds.

The organization of this paper is as follows. In Sec.~\ref{Simulation setup} we explain the MD simulations setup and models for the water and 2,6 lutidine mixture with the   antagonistic salt  PPh$_4$Cl.   Section.~\ref{Results} gives and discusses the simulation results, which is divided into two subsections, concentration profiles \ref{Concentration profiles}, surface tension  and hydrogen bonds \ref{Interfacial tension}. We conclude the paper in Sec.~\ref{Conclusion}. 
\section{Simulation setup}
\label{Simulation setup}
In this work, we consider the same mixture of water/2,6-lutidine provided in  \citep{pousaneh:16:00}. 
For the organic molecule 2,6-lutidine we use the recent parametrisation \citep{pousaneh:16:00} developed using the  Gromacs
package \citep{Spoel:05:00}. The model successfully captures the lutidine bulk properties and the  lutidine/water binary mixture properties, such as the phase diagram, critical properties and  surface tension. We therefore choose the same topology and charge distributions for the lutidine molecule as in ~\citep{pousaneh:16:00}. 
The lutidine molecule C$_7$H$_9$N is presented by 11 atoms with the CH$_3$ groups as  united atoms,  Fig.~\ref{fig:1WEO} (left). For the charge distributions on the lutidine see ~\citep{pousaneh:16:00}.

All simulations in this work are performed using the Gromacs/2016  MD simulation package \citep{Spoel:05:00}.
The Gromos54a7 forcefield \cite{Schmid:11:00} is used for almost all interactions.
Bond lengths are kept fixed at the Gromos54a7 equilibrium length, using the LINCS algorithm \cite{Hess:97:00}.
Water is described using the TIP4P/2005 model.
For the lutidine molecules, the partial charges and dihedrals are taken from   \cite{pousaneh:16:00}.
 The Particle Mesh Ewald (PME) approach is  applied to the electrostatic
interactions  beyond a $1.2$~(nm) cutoff. A cut-off length of $1.2$~(nm) is applied to the    van der Waals  interactions.

Simulations are performed in the NPT ensemble. The
temperature is  controlled by a V-rescale
thermostat at $T=380$~K,  and  the pressure is  controlled by a Parrinello-Rahman barostat (isotropic, $p = 1$~atm). At $T=380$~K  the mixture is in the phase separated region  \cite{pousaneh:16:00}.

\begin{figure}[ht]
\centering 
\includegraphics[width=0.5\textwidth]{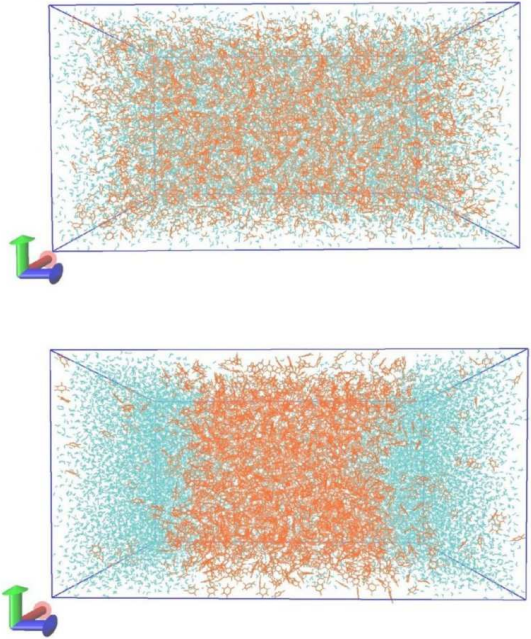}
\caption{(Top) the initial configuration of water/2,6-lutidine box, (bottom) simulation result after $500~$ns at $T=380~$K. Blue color molecules indicate water molecules and the ring orange-coloured indicate lutidine molecules. }
\label{initualbox}\vspace{0.1cm}
\end{figure}

For the binary mixture, we simulate a box containing 19000 water molecules and 3000 lutidine molecules.
This corresponds to $13 \%$ mole fraction of lutidine. 
The initial box size is $(7,7,19) $~nm, see Fig.~\ref{initualbox} (top).
The simulation time step is $2 $~fs.
We run the simulation for $500~$ns to equilibrate the system and allow the water and lutidine to phase separate.
The equilibrated configuration is shown in  Fig.~\ref{initualbox} (bottom).  

Next, we add the antagonistic salt to the mixture. We obtain the topology for the antagonistic salt PPh$_4^+$Cl$^-$  using the Automatic Topology Builder \cite{Malde:11:00} for the  Gromos54a7 forcefield, see Fig.~\ref{fig:1WEO} (right).
 In order to investigate the effect of salt on interfacial tension of water/lutidine mixture, we simulate six different systems with different  salt concentration from $0.045\%$ to $2.70\%$ mole fractions of salt. For every salt ion we add the the box, we remove one water molecule
We then run the simulations for up to $600$ ns depending on the system. Equilibrium typically is reached around $200$-$300$ ~ns. We verify that  the system has reached  equilibrium  not only by checking density, pressure, and temperature, but also the structural properties such as partial densities, radial distribution functions and hydrogen bonds.
\section{Results}
\label{Results}
\begin{figure*}
\includegraphics[width=1\textwidth]{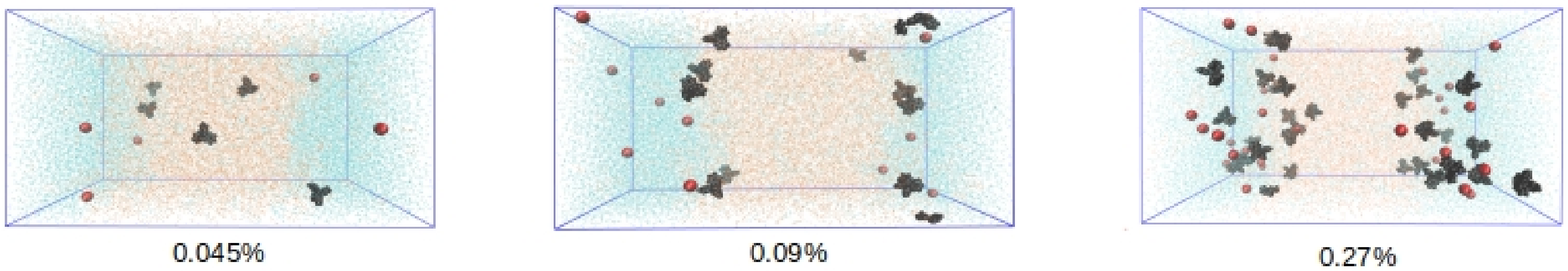}\vspace{-0.2cm}
\includegraphics[width=1\textwidth]{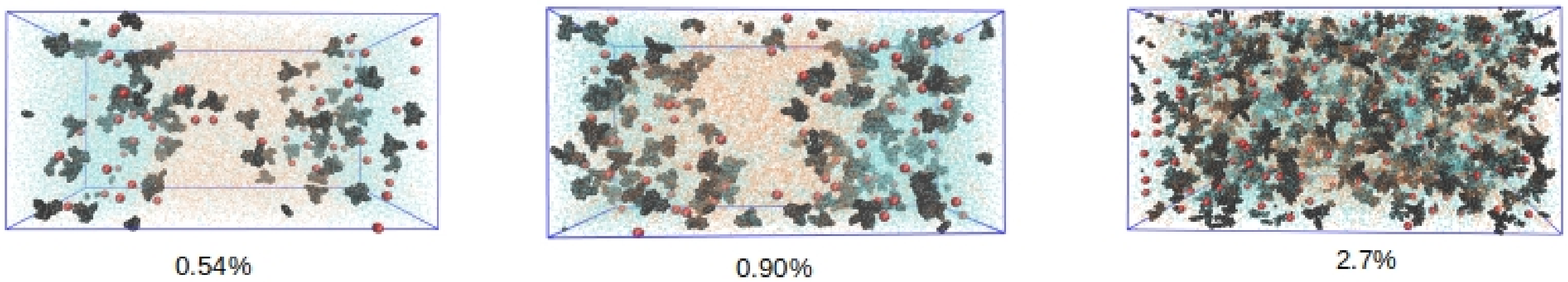}
\vspace{-1cm}
\caption{Equilibrated configurations of the water/2,6-lutidine systems with different salt concentrations.  Red spherical particles indicate the Cl$^-$ anion and black bigger particles indicate PPh$^+$ cations.   The salt concentrations are indicated below each figure.  At lower concentrations, the ions predominantly sits around the interface, but at the highest concentration the interface disappears and the system is mixed. }
\label{config}
\end{figure*}

Examples of equilibrated water/2,6-lutidine/salt configurations are shown in Fig.~\ref{config}.
From the snapshots, it can be seen that the salt tends to  locate around the interface. 
As the concentration increases, the interface becomes saturated and the salt ions are present deeper into the bulk of the water and lutidine.
Finally, at the extremely high concentration of 2.70\% mole fraction the interface disappears completely and the water and lutidine are mixed.
\subsection{Concentration profiles}
\label{Concentration profiles}
\begin{figure}[ht]
\vspace{1cm}
\centering 
\includegraphics[scale=0.3]{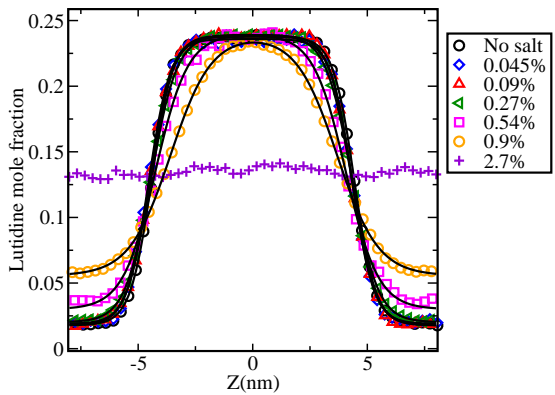}
\caption{Mole fraction of lutidine at different salt concentrations as a function of position in the simulation box along the axis perpendicular to the interfaces.}
\label{fig:lutfraction}\vspace{0.1cm}
\end{figure}

We have investigated the concentration profiles in more detail.
We first focus on lutidine mole fractions in both phases, lutidine-rish and lutidine-poor. Fig.~\ref{fig:lutfraction} shows the lutidine mole fractions as a function of the position in the box, obtained from simulations.
One can see that for higher salt concentrations, the interface between two phases become softer. At the highest concentration, the system becomes homogeneous with water and lutidine mixed. 
The concentration values in the two phases, water-rich and lutidine-rich can be obtained from the classical theories for interfaces between coexisting phases, such as Cahn and Hilliard  or
Landau-Ginzburg theory, which both predict a hyperbolic-tangent shaped
interfacial density profile  \cite{cahn:58:00,pousaneh:16:00,chepela:00:00,Inglesfield:95:00}:
\begin{multline}
w_\mathrm{lut}(z)= w_\mathrm{lut}^\mathrm p+\frac{w_\mathrm{lut}^\mathrm r-w_\mathrm{lut}^\mathrm p}{2} \bigg [  \tanh\bigg(\frac{z-z_0+c}{\lambda}\bigg)-\\
 \tanh \bigg(\frac{z-z_0-c}{\lambda} \bigg) \bigg ],
\label{fiteq}
\end{multline}
with $w_\mathrm{lut}^\mathrm r$ and $w_\mathrm{lut}^\mathrm p$ 
the  fractions  of 2,6-lutidine in the lutidine-rich and lutidine-poor phases respectively.
The width of the interface is given by $\lambda$. $c$ is the half-width of the lutidine-rich region, and $z_0$ is the center of the lutidine-rich phase.
We fit the concentration profiles from the simulations to this function.   These fits are indicated by the solid lines in Fig.~\ref{fig:lutfraction}.
The fit parameters obtained for the lutidine mole fractions in the rich and poor phases are shown in Fig.~\ref{fig:lutfraction-2} and given in the appendix. The  values indicate that  with increasing the salt concentration the lutidine mole fraction in the lutidine-rich phase  decreases, while it increases in the lutidine-poor phase.

\begin{figure}[ht]
\centering 
\includegraphics[scale=0.37]{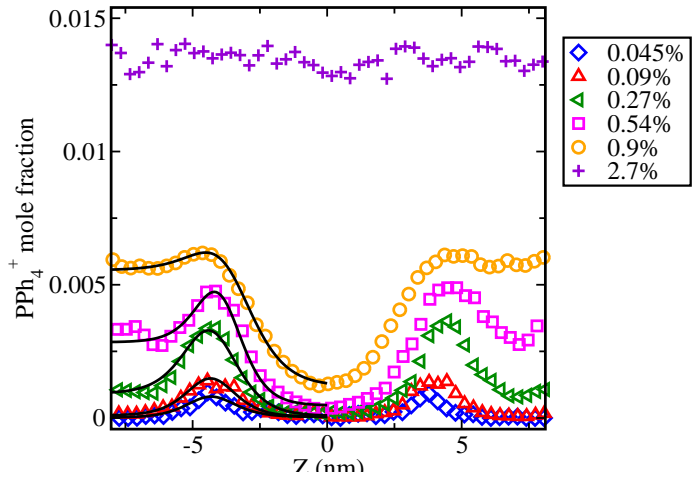}\vspace{1cm}
\hspace{2cm}
\includegraphics[scale=0.37]{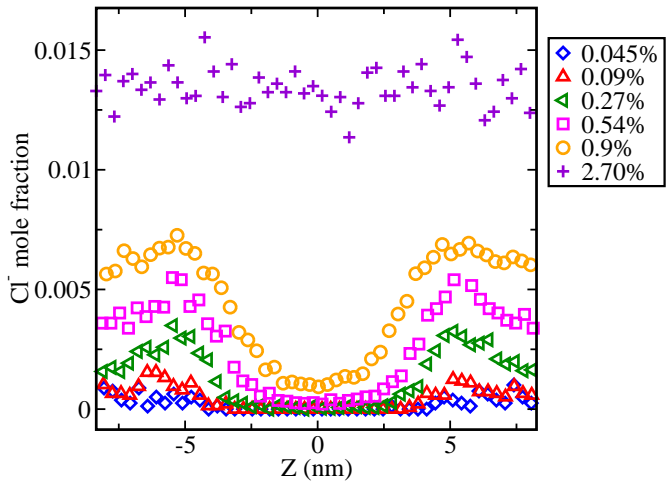}
\caption{The mole fractions of PPh$^+$  (top) and  Cl$^-$ (bottom) for different salt concentrations as a function of position in the simulation box along the axis perpendicular to the interfaces.}
\label{fig:weofraction}\vspace{0.1cm}
\end{figure}

Next, we study the ion concentrations in the two phases and near  the interfaces. Fig.~\ref{fig:weofraction} shows ion concentrations obtained from simulation. The top figure shows the PPh$_4^+$ mole fractions and the bottom one shows the Cl$^-$ mole fractions.
There are clear concentration peaks around the interface between water and lutidine, rather than each ion staying at its preferred phase (hydrophilic anion into water and hydrophobic cations into lutidine).
At extremely high salt concentration, when the water and lutidine are mixed, the salt concentration becomes homogeneous.
In order to obtain the concentration values of  PPh$_4^+$ in each phase, we fit the concentration profiles to the modified hyperbolic-tangent function from Cahn and Hilliard  theory. The modified function  corresponds to an asymmetric interface  where the three  concentrations, at the interface, $w^\mathrm m$, and in the two phases, $w^\mathrm r$,$w^\mathrm l$,  next to the interface, are different.  The function is
\begin{widetext}
\begin{multline}
w(z)=\frac{(w^\mathrm l+w^\mathrm r)}{2}     -\Bigg [ \frac{\tanh(c/\lambda)[w^\mathrm l-w^\mathrm r]+ [w^\mathrm l+w^\mathrm r-2w^\mathrm m]}{4 \tanh(c/\lambda )   } \Bigg ]  \tanh ( \frac{z-z_0+c}{\lambda})  \\
 -\Bigg [ \frac{\tanh(c/\lambda)[w^\mathrm l-w^\mathrm r]-[ w^\mathrm l+w^\mathrm r-2w^\mathrm m]}{4\tanh(c/\lambda)  }   \Bigg ]  \tanh ( \frac{z-z_0-c}{\lambda}) , 
\label{fiteqhalf}
\end{multline}
\end{widetext}
where $w^\mathrm r$,$w^\mathrm l$ and $w^\mathrm m$ are the three concentrations.  
The solid lines in Fig.~\ref{fig:weofraction} (top)  show the fittings to the last equation. They indicate that with increasing salt concentration the amount of ions around the interfaces  increases. This can be better seen in Fig.~\ref{fig:weofraction-2}.  The fit parameters  are  also given in the appendix. 

\begin{figure}[ht]
\centering 
\includegraphics[scale=0.37]{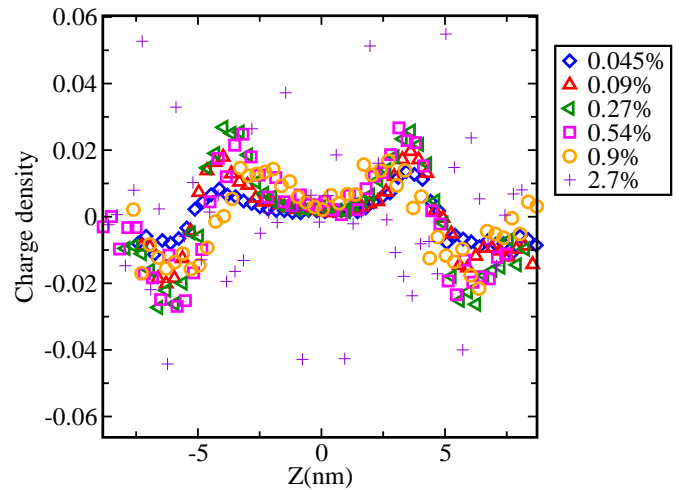}\vspace{0.cm}
\caption{Charge number density  for different salt concentrations as a function of position in the simulation box along the axis perpendicular to the interfaces.}
\label{fig:charge-density}\vspace{0.cm}
\end{figure}

We furthermore show the charge density profile in Fig.~\ref{fig:charge-density}.
The figure illustrates that the positive and negative charges pile up when the salt concentration is increased up to a point.
However, at large salt concentrations they become uniformly distributed  due to  mixing in  the system.
\begin{figure}[ht]
\centering 
\includegraphics[scale=0.3]{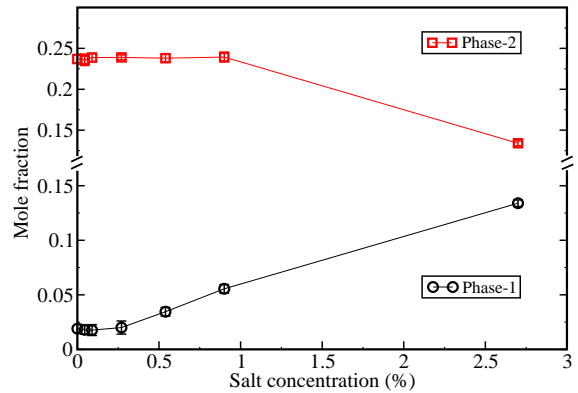}
\caption{Lutidine mole fraction  in two phases, phase-1 (lutidine-poor)  and  phase-2  ( lutidine-rich) obtained from fitting to Eq.~\ref{fiteqhalf}.}
\label{fig:lutfraction-2}\vspace{0.1cm}
\end{figure}
\begin{figure}[ht]
\centering 
\includegraphics[scale=0.3]{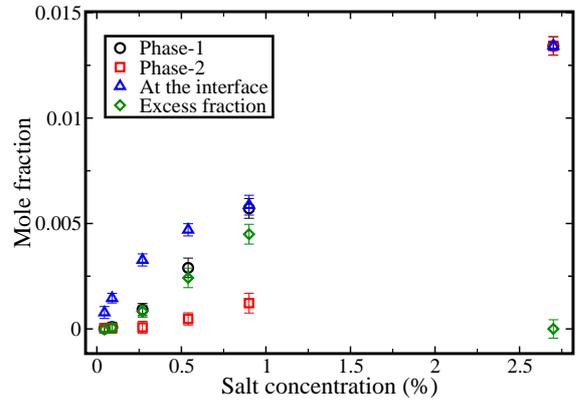}
\caption{PPh$^+$  mole fraction  in different phases, phase-1 (lutidine-poor)  and  phase-2  ( lutidine-rich) and at the interface obtained from fitting to Eq.~\ref{fiteqhalf}. The excess mole fraction of  cation between two phases are aslo shown in green-diamond data. }
\label{fig:weofraction-2}\vspace{0.51cm}
\end{figure}

\subsection{Interfacial tension}
\label{Interfacial tension}
\begin{figure}[ht]
\centering 
\includegraphics[scale=0.3]{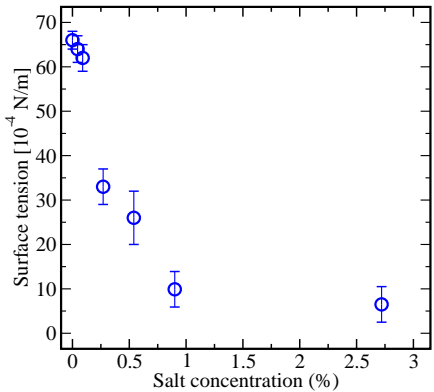}
\caption{Interfacial tension per  interface in  systems with different salt concentration. Interfacial tension decreases by increasing the salt concentration in the system.  
\label{fig:tension}
}
\end{figure}
In order to obtain the interfacial tension we continue  the  simulations in the NPT ensemble  using a Brendsen barostat for $200$~ns. 
In Gromacs, the average interfacial tension $\gamma$ can be calculated directly from the difference between the normal and lateral pressure 
 \begin{eqnarray}
\gamma(t)= \frac{1}{n} \int_0^{L_z} \bigg [P_{zz}(z,t)-\frac{P_{xx}(z,t)+P_{yy}(z,t)}{2}  \bigg ]dz,
\end{eqnarray} 
where $L_z$ is the height of the box and $n$ is the number of surfaces.
The results are  plotted in Fig.~\ref{fig:tension}.
The error estimates for the averages are obtained based on  block averages.
We can see a clear reduction of interfacial tension with increasing salt concentration.

\begin{figure}[ht]
\centering 
\includegraphics[scale=0.25]{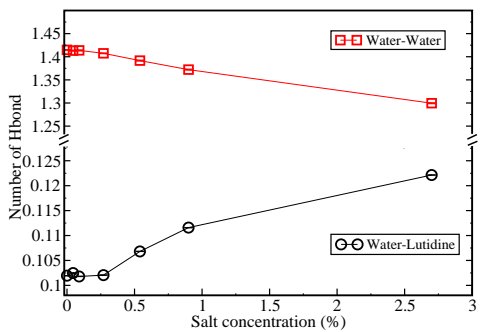}
\caption{Number of hydrogen bonds between water-water and water-lutidine molecules  (per water molecule) in  systems  with different salt concentration. 
\label{fig:HS}
}
\end{figure}

To further investigate the effect of antagonistic salt in the mixture,  we now focus on the hydrogen bonds between lutidine and water.
We obtain the number of hydrogen bonds between water-lutidine molecules, and between water-water molecules  using Gromacs. Hydrogen bonds are determined based on cutoffs for the angle Hydrogen-Donor-Acceptor  and the distance Donor-Acceptor. O and N  are acceptor here \cite{Spoel:05:00}. 
The results are  plotted in Fig.~\ref{fig:HS}  for different salt concentrations.
The number of hydrogen bonds between water and water decreases with increasing salt concentration, while the number of hydrogen bonds between water and lutidine increases. That is a clear indication of the the mixing of the water and lutidine. The error estimate of the averages are obtained based on  block averages.

\section{Conclusion}
\label{Conclusion}
The addition of antagonistic  salt  tetraphenylphosphonium chloride PPh$_4$Cl   in water/2,6-lutidine has been  studied by Molecular Dynamics Simulations.  The salt has   hydrophilic anion and hydrophobic cation, therefore different affinity with the water and lutidine phases. We have examined how the salt effects on the mixture properties. We study the concentration of all four components, we determine  that the ions tend to stay around the interface of the two phases. However, the  ion concentration  profiles in both phases  become the same at a high value of salt concentration. Thus, our simulations reveal that  by increasing the salt concentration  the two-phase region start to shrink  and finally at a high salt concentration,  water and lutidine mixes. 

We obtain surface tension between water and lutidine phases, and show that 
it decreases by increasing the salt concentration. We further investigate the hydrogen bonds between water and lutidine molecules. We report that with increasing the salt concentration, the number of hydrogen bonds between water-water molecules decreases while the number of hydrogen bonds between water-lutidine increases. This can one of indications for the water and lutidine mixing.

It would be interesting to investigate the possibility of mesophases before and
after mixing as well as the structure of the mixed phase  \cite{Tasios:17:00,Araki:09:00}. However,
such a study would necessitate a much bigger system than the one we have
simulated here.  We are limited in the system size by the computational power
required for the detailed atomistic description of the molecules.
\section*{Acknowledgements}
The work has been supported by National Infrastructure for Computational Science in Norway
(UNINETT Sigma2) with computer time for the Center for High Performance Computing (NN9573K and NN9572K). The authors acknowledge The Research Council of Norway for NFR project number 275507   for financial support.  

\section{Appendix. A}
\label{Appendix.A}
The fit parameters obtained for the lutidine  and PPh$_4^+$ mole fractions   in the rich and poor phases are given in Tables.~\ref{tab:lut} and \ref{tab:weo}, respectively. 
\begin{table}
\centering
\vspace{1cm}
\begin{tabular}{|l| c|c|c|c|c|c|}
\hline
   Salt   &Phase-1      & Phase-2  \\
         \hline
         No salt & 0.019 $\pm$ 0.004 &   0.237 $\pm$ 0.008  \\
               \hline
      0.045$\%$ & 0.0179  $\pm$ 0.004&   0.236 $\pm$ 0.007 \\
      \hline
      0.09$\%$ & 0.0177 $\pm$ 0.004 &   0.238 $\pm$ 0.005 \\
      \hline
      0.27$\%$  & 0.020$\pm$ 0.006  &   0.239 $\pm$ 0.003 \\
      \hline
      0.54$\%$ & 0.0345$\pm$ 0.004 &   0.238  $\pm$ 0.005 \\
      \hline
0.90$\%$  & 0.055 $\pm$ 0.004 &   0.239  $\pm$ 0.006 \\
      \hline
      2.70$\%$  & 0.134 $\pm$ 0.003&   0.134 $\pm$ 0.003  \\
      \hline
\end{tabular}
\caption{Lutidine mole fractions   at two phases,  lutidine-poor (phase-1)  and  lutidine-rich (phase-2), for systems with different salt concentrations.   }
\label{tab:lut}
\end{table}

\begin{table}[H]
\vspace{1cm}
\begin{tabular}{|l| c|c|c|c|c|c|c|}
\hline
  Salt    & Phase-1    &Phase-2   & At the interface      \\
           \hline
0.045$\%$  & 0.0014  $\pm$ 0.0004 &   0.0037 $\pm$ 0.0002 & 0.077 $\pm$ 0.003  \\
         \hline
0.09$\%$  & 0.009 $\pm$ 0.0002&   0.004 $\pm$ 0.0002 & 0.14 $\pm$ 0.01\\
      \hline
      0.27$\%$ & 0.091 $\pm$ 0.003 &   0.0076 $\pm$ 0.0003& 0.32 $\pm$ 0.03 \\
      \hline
     0.54$\%$ & 0.28  $\pm$ 0.06&   0.047 $\pm$ 0.003 & 0.46 $\pm$ 0.06 \\
      \hline
          0.90$\%$  & 0.57 $\pm$ 0.08  &   0.122 $\pm$ 0.04 & 0.58$\pm$ 0.04 \\
      \hline
     2.70$\%$  & 1.34 $\pm$ 0.2 &   1.34 $\pm$ 0.2& 1.34 $\pm$ 0.2\\
      \hline
\end{tabular}
\caption{PPh$_4^+$ mole fractions   at two phases, lutidine-poor (phase-1) and lutidine-rich (phase-2), and at the interface, for systems with different salt concentrations. The values are given in unit of $10^{-3}$.  }
\label{tab:weo}
\end{table}

\bibliography{ref}
\bibliographystyle{ieeetr}

\end{document}